\documentclass[intlimits,twoside,a4paper]{article}

\usepackage{amsmath,amssymb}
\usepackage{graphicx}

\usepackage[T2A]{fontenc}
\usepackage[cp1251]{inputenc}

\usepackage[eqsecnum]{cmpj2}

\articletype{Rapid Communication}

\issue{2013}{16}{3}{34601}
\doinumber{10.5488/CMP.16.34601}

\title[Randomly charged polymers]%
{Randomly charged polymers in porous environment}
\author[V. Blavatska, C. von Ferber]{V. Blavatska\refaddr{label1},
        C. von Ferber\refaddr{label2}}
\addresses{
\addr{label1} Institute for Condensed Matter Physics of the National Academy of
Sciences of Ukraine, \\
1 Svientsitskii St., 79011 Lviv, Ukraine
\addr{label2} Applied Mathematics Research Centre, Coventry University, CV1 5FB Coventry,
UK }

\date{Received June 11, 2013, in final form July 11, 2013}
\authorcopyright{V. Blavatska, C. von Ferber, 2013}

\begin{document}

\maketitle

\begin{abstract}
We study the conformational properties of charged polymers in a
solvent in the presence of structural obstacles correlated according
to a power law $\sim x^{-a}$. We work within the continuous
representation of a model of linear chain considered as a random
sequence of charges $q_i=\pm q_0$. Such a model captures the
properties of polyampholytes~-- heteropolymers comprising both
positively and negatively charged monomers. We apply the direct
polymer renormalization scheme and analyze the scaling behavior of
charged polymers up to the first order of an $\epsilon=6-d$,
$\delta=4-a$-expansion.

\pacs  61.25.hp, 11.10.Hi, 64.60.ae, 89.75.Da
\end{abstract}

\section{Introduction}

Many polymer macromolecules encountered in chemical and biological
physics can be considered as long flexible chains. The
conformations of individual macromolecules are in general controlled
by the type of monomer-monomer interactions. In good solvents, where
interactions between monomers are mainly steric, $N$-monomer
homogeneous polymer chains form coil-like structures with the
mean-squared end-to-end distance $R_\textrm{e}$ obeying a scaling law
\cite{deGennes79,desCloizeaux,Grosberg94}:
\begin{equation}
 \langle R_\textrm{e}^2 \rangle  \sim  N^{2\nu_{{\rm coil}}}, \label {RR}
\end{equation}
with a universal exponent depending on space dimension $d$ only [e.g., the phenomenological Flory theory \cite{deGennes79} gives $\nu_\textrm{coil}(d)=3/(d+2)$]. Note that at $d=4$, the intrachain steric
interactions are rendered irrelevant, and the polymer behaves like
an idealized Gaussian chain with $\nu_{{\rm Gauss}}=1/2$.

The long-range nature of the electrostatic Coulomb interaction
between charged monomers produces more complicated effects on
polymer conformations. Of particular interest in this context are
polyampholytes (PAs) \cite{Tanford61} (for a recent review, see e.g.: \cite{Dobrynin04}): heteropolymers
comprising both positively and negatively charged monomers. Examples
of polyampholytes are proteins and synthetic copolymers bearing
acidic and basic repeat groups. The interaction between anionic and
cationic groups leads to additional complications in their physical
behavior. These polymers usually dissolve only when there is
a sufficient amount of  salt added, which screens the interactions between
oppositely charged segments. The properties of polyampholytes  are
successfully captured within the frames of a randomly charged
polymer model \cite{Kantor92a,Kantor92b,Kantor95}: a linear chain composed of a
random sequence of $N$ monomers carrying a charge $\pm q_0$ with
a fixed overall charge $Q$. It is established that the conformational
properties of PAs strongly depend both on $Q$ and the quality of the
solvent (and thus on temperature $T$)
\cite{Edwards80,Kantor92a,Kantor92b,Corpart93,Wittmer93,Gutin94,Kantor95,
Dobrynin96,Barbosa96,Tanaka97,Everaers97a,Everaers97b,Lyubin99,Yamakov00}.
If positive and negative charges are nearly balanced ($Q$ is small),
the attractive Coulomb interaction dominates, and the polymer
collapses into a globular (sphere-like) state, as predicted by the
Debye-H\"uckel theory \cite{Higgs91}. For polyampholytes with
a considerable disbalance between positive and negative charges
($Q>Q_\textrm{c}\sim q_0N^{1/2}$), Coulomb repulsion will be predominant. They
are expected to attain the properties of polyelectrolytes
(homogeneously charged polymers) with a stretched configuration
governed by the size exponent value $\nu\simeq 1$.

In polymer physics,  comprehension of the behavior of macromolecules  in the
presence of structural disorder is of great importance, e.g., in
colloidal solutions \cite{Pusey86}, near microporous membranes
\cite{Cannel80}, or in the crowded  environment of biological cells
\cite{Minton01a,Minton01b,Minton01c}. The density fluctuations of obstacles often lead to
spatial inhomogeneity and create pore spaces of fractal structure
\cite{Dullen79,noninteger1,noninteger2,noninteger3,noninteger4}. In the present study we address a model where the structural obstacles of the environment are
spatially correlated on a mesoscopic scale \cite{Sahimi95}.
Following reference~\cite{Weinrib83}, this case can be described by
assuming the defects to be correlated at large distances $r$
according to a power law with a pair correlation function $g(r)\sim
r^{-a}$.  Such a correlation function describes the defects extended in
space, e.g., the cases $a=d-1$ ($a=d-2$) correspond to lines
(planes) of defects of random orientation, whereas non-integer
values of $a$ can describe obstacles of fractal structures (see
\cite{Weinrib83,Blavatska01a,Blavatska01b,Blavatska01c} for further details). The impact of
long-range-correlated disorder on conformational properties of
polymer chains has been analyzed in previous works
\cite{Blavatska01a,Blavatska01b,Blavatska01c} by means of the field-theoretical
renormalization group (RG) approach. The question how the
characteristics of a polymer with long-range Coulomb interaction are
effected by the presence of such a long-range-correlated
structural disorder  remains, however, still unresolved.

In the present study, we aim to develop a direct renormalization
group approach to analyze the conformational properties of  a random
charge model, proposed in references~\cite{Kantor92a,Kantor92b,Kantor95} in solution
in the presence of long-range-correlated structural obstacles. The
layout of the paper is as follows. In the next section, we develop a
continuous chain representation of the model and discuss its general properties in detail. In section~3, the direct renormalization
procedure is described and the results for the scaling properties of
charged polymers are discussed. We close by giving conclusions
and an outlook.

\section{The model}

Following reference~\cite{Kantor92a,Kantor92b}, we consider a flexible linear chain
consisting of $N$ monomers, each carrying a charge $\pm q_0$. If we
restrict the ensemble of random charged sequences to have a fixed
overall non-zero charge $Q$, correlations within the  positive and
negative charge distributions are induced. Denoting by $\overline
{(\ldots)}$ the average over the distribution of charges along the
chain, one finds~\cite{Kantor95}:
 \begin{eqnarray}
&&\overline {q_i}  = \frac{Q}{N}\,,\qquad \overline {q_i^2}  = q_0^2\,,\qquad \overline{q_i q_j}  = \frac{Q^2-q_0^2N}{N^2}=\frac{Q^2-Q_\textrm{c}^2}{N^2}\, ,\qquad i\neq j\,,
\label{cor}
 \end{eqnarray}
with $Q_\textrm{c}=\sqrt {\phantom{5}\sum_{i=1}^N\overline{\,q_i^2}\phantom{5}}=q_0N^{1/2}$.

Let us pass to a continuous model, where the polymer chain in a porous
environment is represented by a path $r(s)$, parameterized by $0\leqslant
s\leqslant L$. The probability of each path configuration is given by the
Boltzmann distribution function:
\begin{eqnarray}
P\left(\{\vec{r}\}\right)&=& \exp\left\{-\frac{1}{2}\int_0^{L}{\rm d}s
\left[\frac{{\rm d} r(s)}{{\rm d} s}\right]^2-\frac{u_0}{2}
\int_0^{L}{\rm d}s'\int_0^{L}{\rm d}s{''}\,\delta\big[{\vec{r}}(s')-{\vec{r}}(s{''})\big]\right.\nonumber\\
&&-\left. \frac{1}{2} \int_0^{L}{\rm d}s' \int_0^{L} {\rm d}s''
\frac{q(s')q(s'')}{\left|\vec{r}(s')-\vec{r}(s'')\right|^{d-2}}-\int_0^{L}{\rm d}s\, V\big[\vec{r}(s)\big]\right\}. \label{Prob}
 \end{eqnarray}
Here, the first term in the exponent represents the chain
connectivity, the second term describes the short range excluded
volume interaction with the coupling constant $u_0$, the third term
gives the unscreened electrostatic interaction in $d$ dimensions
(Coulomb potential), where the function $q(s)$ represents the
charges along the chain in a particular configuration, and the last
term arises due to the steric interactions between the polymer chain
and the structural defects in the environment given by the potential
$V\big[\vec{r}(s)\big]$. Following \cite{Weinrib83}, we assume the pair
correlation function of defects to decay with distance according to
the scaling law:
\begin{eqnarray}
\left\langle V\big[\vec{r}(s')\big]V\big[\vec{r}(s'')\big]\right\rangle = w_0|\vec{r}(s')-\vec{r}(s'')|^{-a},\label{corr}
\end{eqnarray}
here, $\langle \ldots \rangle$ denotes the average over different realizations of the disordered
environment).

Taking into account that only the two last terms in (\ref{Prob})
include random variables,  we find the averaged partition function $
\overline{ \langle Z(L) \rangle}= \int {\cal D}r \exp(-{\cal
H}_\textrm{eff}) $ with an effective Hamiltonian:
\begin{eqnarray}
{\cal H}_\textrm{eff}&=&\frac{1}{2}\int_0^{L}{\rm d}s
\left[\frac{{\rm d} r(s)}{{\rm d} s}\right]^2+\frac{u_0}{2}
\int_0^{L}{\rm d}s'\int_0^{L}{\rm d}\,s{''}\,\delta\big[{\vec{r}}(s')-{\vec{r}}(s{''})\big]\nonumber\\
&&+\frac{v_0}{2}\int_0^{L}{\rm d}s' \int_0^{L} {\rm d}s''
\frac{1}{\left|\vec{r}(s')-\vec{r}(s'')\right|^{d-2}}+\frac{w_0}{2}\int_0^{L}{\rm d}s' \int_0^{L} {\rm d}s''
|\vec{r}(s')-\vec{r}(s'')|^{-a}. \label{Hef}
\end{eqnarray}
Here, we introduce the notation $v_0\equiv{\overline{q(s)q(s')}}\sim
(Q^2-Q_\textrm{c}^2)$.  We  thus have a model with three types of intrachain
interactions governed by coupling constants  $u_0$,  $v_0$ and
$w_0$. From a dimensional analysis of the couplings
$ [u_0]=L^{-(4-d)/2}$, $[v_0]=L^{-(6-d)/2}$, $[w_0]=L^{-(4-a)/2}
$
one easily concludes that for Coulomb interaction, the upper
critical dimension $d_\textrm{c}=6$, whereas for the excluded volume
interaction $d_\textrm{c}=4$, and thus the latter may be neglected in the
renormalization group scheme \cite{Pfeuty77,Jug81,Amit}. In what
follows, we thus restrict our considerations to a model with
only two couplings $v_0$ and $w_0$. The sign of the coupling $v_0$
depends on the overall charge $Q$ of a given polyampholyte: a
positive value of $v_0$ (i.e., $Q^2>Q_\textrm{c}$) corresponds to the case,
where the Coulomb repulsion is predominant in determining the
conformation of the polymer chain (and thus the polymer is expected
to expand).

Note that in deriving equation~(\ref{Hef}) we restrict the consideration  to a simpler case of annealed disorder averaging, taking into account that for
an infinitely long single polymer chain in random disorder, the distinction between quenched and annealed averages is negligible~\cite{Cherayil90a,Cherayil90b,Cherayil90c,Cherayil90d}.

\section{Renormalization and results}

We follow the direct polymer renormalization scheme, developed by
des Cloizeaux \cite{desCloizeaux}, generalizing it to the case of
two intrachain interactions. Working within the frames of the
continuous polymer chain model, one encounters problems with various
divergences as the polymer length diverges (and thus
the number of configurations tends to infinity). However, all
divergencies can be eliminated by introducing renormalization
factors, allowing to define and directly estimate the physical
quantities of interest. The renormalization procedure is related to
the existence of universal critical indices and critical factors.

Expanding the theory in the couplings $v_0$ and $w_0$ and passing to
the Fourier transform according to
\[
\left|\vec{r}(s')-\vec{r}(s'')\right|^{-a}\simeq\int{\rm d}
\vec{k}\,|k|^{a-d}\exp\left\{\ri\vec{k}\left[{\vec{r}}(s')-{\vec{r}}(s{''})\right]\right\},
\]
we find for the partition function of the system:
\begin{eqnarray}
&&\overline{\langle Z(L)\rangle}={Z^0}(L)\left[1{-}\frac{4\,v_0(2\pi)^{-\frac{(d-2)}{2}}L^{3-\frac{d}{2}}}{\left(4-d\right)\left(6-d\right)}-\frac{4\,w_0(2\pi)^{-\frac{a}{2}}L^{2-\frac{a}{2}}}{\left(2-a\right)
\left(4-a\right)} \right].
\label{zzlast}
\end{eqnarray}
Here,
\[
{Z^0}(L)=\int {\cal D}r\,\exp\left\{{-\frac{1}{2}\int_0^{L}{\rm
d}s \left[\frac{{\rm d} r(s)}{{\rm d} s}\right]^2}\right\}
\]
denotes the
partition sum of the unperturbed model (Gaussian chain). Note that
only the first order terms are kept in the above relation (we
restrict ourselves to the so-called ``one-loop approximation'' of
perturbation theory).

The averaged squared end-to-end distance $R_\textrm{e}^2=[r(L)-r(0)]^2$  of a
typical polymer chain configuration may be calculated using the
identity:
\begin{eqnarray}
\label{ras} (\overline{\langle R^2_e \rangle})_{\cal H} =
\left\{-2d\frac{\partial}{\partial q^2}{\rm
e}^{\ri\vec{q}\,\left[r(L)-r(0)\right]^2}\right\}_{\cal H}
\end{eqnarray}
with:
\begin{equation}
(\ldots)_{\cal H}= \frac{\int {\cal D}r\, (\ldots) \, {\rm
e}^{-{\cal H}_\textrm{eff}}}{\overline{\langle Z(L) \rangle}}\,. \label{aver}
\end{equation}
For the one-loop approximation we find:
\begin{equation}
(\overline{\langle R_\textrm{e}^2\rangle})_{\cal H}
{=}Ld\left[1{+}\frac{4\,v_0(2\pi)^{-\frac{d-2}{2}}L^{3-\frac{d}{2}}}{\left(6-d\right)\left(8-d\right)}{+}
\frac{4\,w_0(2\pi)^{-\frac{a}{2}}L^{2-\frac{a}{2}}}{\left(4-a\right)\left(6-a\right)}
\right]. \label{rr}
\end{equation}
We may thus define a new (renormalized) scale $L_\textrm{R}$ and introduce a swelling factor $\chi_2$ via: $(\overline{\langle R_\textrm{e}^2\rangle})_{\cal H}=dL_\textrm{R}$, $\chi_2(v_0,w_0)={L_\textrm{R}}/{L}$.
Remembering that $(\overline{\langle R_\textrm{e}^2\rangle})_{\cal H}\sim L^{2\nu}$ and thus $\chi_2\sim L^{2\nu-1}$ one has:
\begin{equation}
(2\nu-1)=L\frac{\partial \chi_2(v_0,w_0)}{\partial L}\,.\label{nuu}
\end{equation}

The final renormalization step can be performed by analyzing the
virial expansion  for the osmotic pressure of a solution of polymers
\cite{desCloizeaux}. To this end, we need the contributions to the
partition functions ${\overline{\langle
Z}(L,L)\rangle}=\overline{\langle
Z_{v_0}(L,L)\rangle}+\overline{\langle Z_{w_0}(L,L)\rangle}$ of the
system of two interacting polymer chains of the same length $L$. The
dimensionless renormalized coupling constants $v$, $w$ are thus
defined by:
\begin{eqnarray}
&&v=-\frac{\overline{\langle Z_{v_0}(L,L)\rangle}}{\overline{\langle Z^{2}(L)\rangle}}L_\textrm{R}^{-\frac{d-2}{2}},\qquad w=-\frac{\overline{\langle Z_{w_0}(L,L)\rangle}}{\overline{\langle Z^{2}(L)\rangle}}L_\textrm{R}^{-\frac{a}{2}}.
\end{eqnarray}
The RG flows of renormalized coupling constants are governed by
the functions: $ \beta_u=L{\partial \ln v}/{\partial
L}$, $\beta_w=L{\partial \ln w}/{\partial L}. $ The
corresponding expressions read:
\begin{eqnarray}
&&\beta_v=(6-d)v-\frac{2v^2(d-2)}{(8-d)}-\frac{2vw(d-2)}{6-a}=\epsilon v-4v^2-4vw,\nonumber\\
&&\beta_w=(6-a)w-\frac{2w^2a}{(6-a)}-\frac{2vw a}{8-d}=\delta w-4w^2-4vw.\label{wwfinal}
\end{eqnarray}
Above, we have performed a double $\epsilon=6-d$, $\delta=4-a$
expansion, keeping terms up to linear order in these parameters. The
fixed points $u^*$, $v^*$ of the renormalization group
transformations are defined as  common zeros of the RG functions
(\ref{wwfinal}). We find three distinct fixed points determining
the scaling behavior of a system at different $a$ and $d$:
\begin{align}
&\text{Gaussian:}  & &v^*=0,& & w^*=0, & &\text{stable for}& &  \varepsilon,\delta<0,\\
&\text{Coulomb:} & & v^*=\frac{\epsilon}{4}, & &  w^*=0, & & \text{stable for} & &\delta<\epsilon,\\
&\text{\rm Disorder:} & &v^*=0, & &
w^*={\delta}/{4},& & \text{stable for} & &\delta>\epsilon.
\end{align}
Let us analyze the above results  more in detail. The Gaussian fixed
point corresponds to the situation, where any monomer-monomer
interaction is irrelevant. This happens when we are above the upper
critical dimensions of all interactions ($d>6$, $a>4$).  In the case
when $a>d-2$, the presence of correlated defects plays no role, and
the Coulomb fixed point is stable (only the electrostatic
interaction is relevant in this case). The fixed point obtained has
a positive value, and thus we restore the behavior for $Q>Q_\textrm{c}$
(polyelectrolyte limit). Finally, in the case where $a<d-2$, the
strongly correlated disorder causes the main effect on the polymer
behavior, and the Coulomb interaction is now irrelevant (the
Disorder fixed point is stable).
The absence of a stable fixed point where both long range interactions are present
can be explained using dimensional analysis [see explanation after (\ref{Hef})]: at $a>d-2$, coupling $w_0$ becomes dimensionless
for $d<6$, and thus it is irrelevant in the renormalization group sense,
whereas at $a<d-2$ it is dimensionless for $d>6$ and thus
the interaction $v_0$ is irrelevant.
The critical exponents $\nu$, governing the size measure of a
charged polymer in all three situations, described above, are found
by evaluating (\ref{nuu}), rewriting the corresponding expressions in
terms of renormalized couplings and finally substituting the values
of fixed points listed above. We find:
\begin{align}
\nu_{{\rm Gauss}}\ \ \ &=\frac{1}{2}\,,\\
\nu_{{\rm Coulomb}}&=\frac{1}{2}+\frac{\epsilon}{8}\,,\\
\nu_{{\rm Disorder}}\,&=\frac{1}{2}+\frac{\delta}{8}\,. \label{nuLR}
\end{align}
Note that $\nu_{{\rm Coulomb}}$ coincides with the critical exponent
governing the stretching of a polyelectrolyte chain estimated
previously within the RG scheme in  references~\cite{Pfeuty77,Jug81}.
Thus, we conclude that PAs with any $Q>Q_\textrm{c}$ belong to the
universality class of polyelectrolyts. For the physically
interesting case $d=3$ ($\varepsilon=3$) we may estimate
$\nu_{{\rm Coulomb}}\simeq 0.88$. According to (\ref{nuLR}), any neutral
chain in solution in the presence of long-range-correlated defects,
governed by correlation function with $a=d-2$, obeys exactly the
same scaling behaviour as a charged polymer with unscreened Coulomb
interaction in a pure solvent.

\section{Conclusions}

We studied the scaling properties of charged polymers
(polyampholytes) in a solvent in the presence of structural
obstacles spatially correlated on a mesoscopic scale according to a
power law $\sim x^{-a}$ \cite{Weinrib83}. Such correlations can
describe extended pore-like defects of fractal structure. Within the
randomly charged polymer model,  a polyampholyte is considered as a
linear chain composed of a random sequence of $N$ monomers, each
carrying a charge $\pm q_0$ \cite{Kantor92a,Kantor92b,Kantor95}. The model predicts a contraction of the
charged polymer size when the total charge $Q<Q_\textrm{c}$ and an expansion
for $Q>Q_\textrm{c}$ (with $Q_\textrm{c}\sim N^{1/2}$). The presence of long-range
correlated disorder leads to additional steric interaction between
monomers in addition to the long-range Coulomb potential.

Passing to the  continuous chain limit with two types of
interactions (electrostatic and steric), we developed a direct
renormalization group approach by generalizing the scheme of des
Cloizeaux \cite{desCloizeaux} and analyzed the peculiarities of
conformational transitions, which can be observed in charged
polymers in a disordered environment. In the case where $a>d-2$, the
presence of correlated defects plays no role, and the Coulomb fixed
point is stable (only electrostatic interaction is relevant in this
case). The fixed point obtained has a positive value, and thus we
restore the behavior at $Q>Q_\textrm{c}$ (polyelectrolyte limit
\cite{Pfeuty77,Jug81}). Thus, PAs with any $Q>Q_\textrm{c}$ belong to the
universality class of  polyelectrolytes. At $a<d-2$, the strongly
correlated disorder causes the main effect on the polymer behavior,
while Coulomb interaction is irrelevant.  In particular, one may
conclude that any neutral chain in solution in the presence of
long-range-correlated defects, governed by correlation function with
$a=d-2$, follows exactly the same scaling behaviour, as a charged
polymer with unscreened Coulomb  interaction in a pure solvent.

\section*{Acknowledgement}
This work was supported in part by the FP7 EU IRSES project N269139
``Dynamics and Cooperative Phenomena in complex Physical and
Biological Media''.

\ukrainianpart
\title{Випадково заряджені полімери в пористому середовищі}
\author{В. Блавацька\refaddr{label1}, К. фон Фербер\refaddr{label2}}
\addresses{
\addr{label1} Інститут фізики конденсованих систем НАН України,
вул. Свєнціцького, 1, 79011 Львів, Україна
\addr{label2} Дослідницький центр прикладної математики, Університет Ковентрі, CV1 5FB Ковентрі, Англія
}

\makeukrtitle

\begin{abstract}
\tolerance=3000%
Досліджуються конформаційні властивості заряджених полімерів в розчині у присутності структурних неоднорідностей,
скорельованих згідно степеневого закону $\sim x^{-a}$.
Використовується модель, в якій полімерний ланцюжок представлено як випадкову послідовність зарядів $q_i=\pm q_0$. Така модель
описує властивості поліамфолітів -- гетерополімерів, що містять як позитивно, так і негативно заряджені групи мономерів.
Застосовується підхід прямого полімерного перенормування і аналізується скейлінгова поведінка заряджених полімерів
до першого порядку подвійного $\epsilon=6-d$, $\delta=4-a$-розкладу.
\keywords полімери, заморожений безлад, скейлінг, ренормалізаційна група
\end{abstract}

\end{document}